\documentclass[10pt]{revtex4}
\pdfoutput=1
\usepackage{amssymb}
\usepackage{latexsym}
\usepackage{epsfig}
\begin{document}

\title{ Teleparallel loop quantum cosmology in a system of intersecting branes }

\author{Alireza Sepehri $^{1,2}$}
\email{alireza.sepehri@uk.ac.ir} \affiliation{ $^{1}$Faculty of
Physics, Shahid Bahonar University, P.O. Box 76175, Kerman,
Iran.\\$^{2}$ Research Institute for Astronomy and Astrophysics of
Maragha (RIAAM), P.O. Box 55134-441, Maragha, Iran. }

\author{Anirudh Pradhan $^{3}$}
\email{pradhan.anirudh@gmail.com} \affiliation{ $^{3}$ Department of Mathematics, Institute of
Applied Sciences and Humanities, G L A University, Mathura-281 406, Uttar Pradesh, India. }

\author{A. Beesham $^{4}$}
\email{beeshama@unizulu.ac.za} \affiliation{ $^{4}$ Department of Mathematical Sciences, University of
Zululand, Kwa-Dlangezwa 3886, South Africa.}

\author{Jaume de Haro$^{5}$}\email{jaime.haro@upc.edu}
\affiliation{ $^{5}$ Departament de Matem\`atiques, Universitat Polit\`ecnica de Catalunya, 
Diagonal 647, 08028 Barcelona, Spain}

\begin{abstract}
Recently, some authors have removed the big bang singularity in teleparallel
Loop Quantum Cosmology (LQC) and have shown that the universe may undergo a
number of oscillations. We investigate the origin of this type of teleparallel
theory in a system of intersecting branes in $M$-theory in which the angle
between them changes with time. This system is constructed by two intersecting
anti-$D8$-branes, one compacted $D4$-brane and  a $D3$-brane. These branes
are built by joining $M0$-branes which develop in decaying fundamental strings.
The compacted $D4$-brane is located between two intersecting anti-$D8$ branes
and glues to one of them. Our universe is located on the $D3$ brane which wraps around the $D4$ brane
from one end and sticks to one of the anti-$D8$ branes from the other one. In this system,
there are three types of fields, corresponding to compacted $D4$ branes, intersecting branes
and $D3$-branes. These fields interact with each other and make the angle between
branes oscillate. By decreasing this angle, the intersecting anti-$D8$ branes approach 
 each other, the $D4$ brane rolls, the $D3$ brane wraps around the $D4$ brane, and 
the universe contracts. By separating the intersecting branes and increasing the angle, the $D4$
brane rolls in the opposite direction, the $D3$ brane separates from it and the expansion branch
begins. Also, the interaction between branes in this system gives us the exact form of the relevant
Lagrangian for teleparallel LQC. \\\\


 \end{abstract}
 \date{\today}

\maketitle
\section{Introduction}
Until now, many scientists have tried to propose a model which removes the big bang and predicts
phenomenological events. One of the best theories which prevents this singularity by introducing a
modification in Friedmann's equation is  Loop Quantum Cosmology (LQC). The holonomy corrections
in the flat Friedmann-Lema{\^\i}tre-Robertson-Walker (FLRW) geometry can be
added to the classical Hamiltonian by replacing  the Ashtekar connection $\bar{c}{{}\equiv \gamma \dot{a}}$, where $\gamma$ is
the Barbero-Immirzi parameter, by the function $\frac{\sin(2\bar{\mu}\bar{c})}{2\bar{\mu }}$, where
$\bar{\mu}= \frac{3\sqrt{3}}{2}$ (see for instance \cite{R1a}). Then,
with the help of this new holonomy corrected Hamiltonian, one derives the modified Friedman equation
(an ellipse in the plane ($H,\rho$), where $H$ is the Hubble parameter and $\rho$ the energy density of the universe) \cite{R1}. The same holonomy corrected Friedmann equation can be provided  in the context of teleparallel gravity, considering a 
$F({\mathcal T})$-Lagrangian density -named as teleparallel LQC - where in the flat FLRW spacetime the scalar torsion is given by ${\mathcal T}=-6H^{2}$ \cite{R2}. From the viewpoint 
 of teleparallel LQC, the universe evolves from the contracting phase to the expanding one passing through a non-singular bounce \cite{R3}. Moreover, this theory mixes the simple bounce obtained by holonomy
corrections in LQC with the non-singular perturbation equations given by $F({\mathcal T})$ gravity and derives 
a non-singular bounce scenario as a viable alternative to the inflationary paradigm \cite{R3a,R4}. In parallel,
there are some models in string theory which replace the big bang singularity by the fundamental string
and predict that the age of the universe is infinity \cite{R5,R6,R7,R8,R9,R10,R11,R12}. In these models,
firstly, $N$ fundamental strings are excited and decay to $N$ pairs of $D0$-anti-$D0$-branes in string theory
or $M0$-anti-$M0$-branes in $M$-theory. Then, these branes stick to each other and construct a universe and
an anti-universe in additional to one wormhole. This wormhole is a channel for flowing energy from extra
dimensions into our universe and causes the evolution of the universe from  inflation to late time acceleration. \\

Now, the question arises as to what is the relation between teleparallel LQC and cosmological models in string
theory? We answer this question with a system of two intersecting anti-$D8$-branes in which the angle between them
changes with time. In this system, a compacted $D4$-brane is located between two intersecting branes and glues to
one of the anti-$D8$-branes. The $D3$ brane sticks to  one of the anti-$D8$-branes from one end and wraps around
the $D4$ brane from the other end. Also, there are three types of fields which live on the $D3$, anti-$D8$ and
$D4$ branes and lead to oscillation of the angle between the intersecting branes. By decreasing the angle, two
intersecting branes come close to each other, the $D4$ brane rolls, the $D3$ brane wraps around the $D4$ brane
and the  universe contracts. By increasing the angle, the two intersecting branes move apart from each other, the
$D4$ brane rolls in the opposite direction, the $D3$ brane opens from it and the universe expands. This model
gives us the exact form of the relevant action in teleparallel LQC.\\

The outline of the paper is as follows. In section \ref{o1}, we will construct teleparallel LQC in a system of
intersecting branes and obtain the relevant Lagrangian. In section \ref{o2}, we will consider the origin of
torsion in Einstein cosmology. The last section is devoted to a summary and conclusion. \\

The units used throughout the paper are: $\hbar=c=8\pi G=1$.

\section{Teleparallel loop quantum cosmology in a system of intersecting branes }\label{o1}
In this section, we will show that firstly, $N$ fundamental strings decay to $N$ pairs of $M0$-anti-$M0$-branes.
Then these branes join to one another and form two intersecting anti-$D8$ branes, a compacted $D4$-brane and a
$D3$-brane. We will show that the oscillation of the angle between the anti-$D8$-branes leads to the emergence of
teleparallel LQC and we re-obtain its Lagrangian in this system. \\

Firstly, let us introduce the relevant Lagrangian of teleparallel LQC \cite{R3,R3a,R4}:
\begin{eqnarray}
&& L=V F({\mathcal T})-V\rho \label{t1} \; ,
\end{eqnarray}
where $V=a^{3}$ is the volume,  $a$ being the scale factor of universe. Also, when the universe is filled by a barotropic fluid, the energy density of the universe $\rho$ has to be understood as a function of the volume $V$,  and the Hubble parameter  can be expressed as $H=\frac{1}{3V}\frac{dV}{dt}$. On the other hand , the gravitational part of the Lagrangian is given by
\begin{eqnarray}
&&F_{\pm}({\mathcal T})=\pm \sqrt{-\frac{{\mathcal T} \rho_{c}}{2}} \arcsin \left(\sqrt{-\frac{{{}2}{\mathcal T}}{{{} \rho_{c}}}}\right)
{{} + \frac{\rho_{c}}{2}\left(1\pm\sqrt{1+\frac{2{\mathcal T}}{\rho_{c}}}\right)} \; ,
\end{eqnarray}
where ${\mathcal T}=-6 H^{2}$ is the scalar torsion in the flat FLRW space-time and $\rho_{c}$ is the critical energy density (the maximum value of the energy density), i.e., the energy density of the universe at the bouncing time. \\

Then, the Hamiltonian constraint in teleparallel gravity
\begin{eqnarray}
{\mathcal H}\equiv \dot{V}\partial_{\dot{V}}L-L=(2TF'({\mathcal T})-F({\mathcal T})+\rho)V=0
\end{eqnarray}
leads to the modified Friedmann equation
\begin{eqnarray}
H^2=\frac{\rho}{3}\left(1-\frac{\rho}{\rho_c} \right)\Longleftrightarrow
\rho=G_{\pm}=\frac{\rho_{c}}{2}\left(1\pm\sqrt{1+\frac{2{\mathcal T}}{\rho_{c}}}\right) \; ,
\label{t2}
\end{eqnarray}
which depicts an ellipse in the plane $(H,\rho)$. \\

We will construct this Lagrangian in a system of intersecting branes and show that the universe oscillates
by increasing and decreasing the angle between branes such as $(F_{+},G_{+})$ which correspond to the upper
branch of the ellipse and $(F_{-},G_{-})$ which is the lower one. In our system, there are three fields
that live on the anti-$D8$, $D4$ and $D3$ branes, named, respectively,  $A,$  $\Theta$ and $E$. We will
show that by approaching the intersecting branes, these fields gain negative mass and transit to a tachyon $T$.
In these conditions, the tachyonic  action of the $D3$ brane can be written as \cite{R13}:
\begin{eqnarray}
&& S_{DBI-D3} =-T_{3} \int {d^{4}\sigma} ~ V(T)\sqrt{-det D}\label{t3} \; ,
\end{eqnarray}
where
\begin{eqnarray}
&&D_{ab}=P\left[g_{ab}-\frac{T^{2}}{2\pi
\alpha'Q}g_{ai}l^{i}l^{j}g_{bj}\right]+2\pi \alpha' F_{ab}\nonumber
\\&& + \frac{1}{Q}\Big(\pi\alpha'(D_{a}T(D_{b}T)^{\ast})+D_{b}T(D_{a}T)^{\ast})\Big)\nonumber
\\&& + \frac{i}{2}(g_{ai}+\partial_{a}X^{j}g_{ji})l^{i}(T(D_{b}T)^{\ast}-T^{\ast}(D_{b}T))\nonumber
\\&& +\frac{i}{2}(T(D_{a}T)^{\ast}-T^{\ast}(D_{a}T))l^{i}(g_{bi}+\partial_{b}X^{j}g_{ji}) \; , \nonumber
\\&&\nonumber
\\&& Q=1+\frac{T^{2}}{2\pi \alpha'}l^{i}l^{j}g_{ij}\quad D_{a}T=\partial_{a}T-i E_{a}T \; ,
\nonumber
\\&& V(T)=\frac{1}{{{}\cosh(\sqrt{\pi T})}}\quad F_{ab}^{k}=\varepsilon^{ijk}E_{a}^{i}E_{b}^{j}\label{t4} \; .
\end{eqnarray}

Here, $g_{ij}$ is the metric of the D3-brane and $E_{b}^{j}$ is the field that lives on the D3-brane. This field has a direct relation with the metric of the D3-brane ($\langle E_{a},E_{b}\rangle=g_{ab}$). Our universe is located on the D3-brane and thus these fields can be written in terms of the scale factor of the  universe  ($E^{a}=(-1.a,a,a),\quad E_{a}=(1,a,a,a),\quad g^{ab}=(-1,a^{2},a^{2},a^{2})=F^{ab}$ where a is the scale factor of the universe and F is the field strength on the brane). Also, $T_{3}$ is the tension of the $D3$-brane, $T$ is the tachyon, $\alpha'$ is the string coupling, $a,b$ are
related to the tangent directions of the $D$-branes, while $i,j$ correspond to the transverse one. On the other hand,
$l^{i}$ is the separation between  two anti-$D8$-branes, $P=\eta_{MN}\partial_{a}X^{M}\partial_{b}X^{N}$ and
$M,N=0,1,2,...,9$. The  world-volume coordinates of the $D3$-brane are $(x_{0},x_{1},x_{2},x_{3})$, $l^{i}=r \Theta$
and $r$ is the length of the anti-$D8$-brane. For simplicity, we assume that that all fields
depend only on time and obtain the relevant tachyonic action of the $D3$ brane \cite{R13}:

\begin{eqnarray}
&&S_{DBI-D3}=\frac{-2T_{3}}{g_{s}}\int {d^{4}} ~ V(T)\sqrt{1+\frac{r^{2}}{4}\Theta'^{2}+2\pi \alpha'
T'^{2}+\frac{r^{2}}{2\pi \alpha'}\Theta^{2}T^{2}+F_{ab}^{2}} \; .
\label{t5}
\end{eqnarray}
Assuming that the fields are very small and $V(T)=1$, we can rewrite the above action as follows:
\begin{eqnarray}
&&S_{DBI-D3} = -\frac{2T_{3}}{g_{s}}\int{d^{4}}\left(1+\frac{r^{2}}{4}\Theta'^{2}+2\pi \alpha'
T'^{2}+\frac{r^{2}}{2\pi\alpha'}\Theta^{2}T^{2}+F_{ab}^{2}\right) \; .
\label{tt5}
\end{eqnarray}
When intersecting branes move away from each other, the tachyon is
replaced by the usual fields which live on the $D8$ brane ($ A^{i}$). In fact,
by approaching branes and anti-branes towards each other, these
fields gain negative mass and transit to tachyons. By rebounding
branes, these fields return to their usual state. Under these conditions,
the action (\ref{tt5}) can be rewritten as:

\begin{eqnarray}
&&S_{DBI-D3}=\frac{-2T_{3}}{g_{s}}\int d^{4}x
\left(1+\frac{r^{2}}{4}\Theta'^{2}+2\pi \alpha' \left(\frac{d
A^{i}}{dt}\right)^{2}+\frac{r^{2}}{2\pi
\alpha'}\Theta^{2}(A^{i})^{2}+F_{ab}^{2}\right) \; .
\label{ttt5}
\end{eqnarray}

We can show that this action is built by summing over the actions
of $M0$-branes. To do this, we will use the method \cite{R5}.
In that mechanism,  a fundamental string  is excited and  transits
to a pair of $M0$-anti-$M0$-branes in addition to some extra energy
($V$) \cite{R5}:

\begin{eqnarray}
&&  S_{F-string} = S_{M0}+S_{anti-M0}+2V(extra) \; .
\label{t6}
\end{eqnarray}
Here, $V_{Extra}= -6T_{M0}\int dt\Sigma_{M,N,L,E,F,G=0}^{9}\varepsilon_{MNLD}\varepsilon_
{EFG}^{D}X^{M}X^{N}X^{L}X^{E}X^{F}X^{G}$ and the action of the $M0$-branes is given by
\cite{R5,R14,R15,R16,R17,R18,R19,R20,R21,R22,R23}:

\begin{eqnarray}
&&S_{M0} = S_{anti-M0}  =T_{M0}\int dt Tr( \Sigma_{M,N,L=0}^{10}
\langle[X^{M},X^{N},X^{L}],[X^{M},X^{N},X^{L}]\rangle) \; ,
\label{t7}
\end{eqnarray}
where  $T_{M0}$ is the brane tension, $X^{m}$ are transverse
scalars,  $X^{M}=X^{M}_{\alpha}T^{\alpha}$, and

\begin{eqnarray}
 &&[T^{\alpha}, T^{\beta}, T^{\gamma}]= f^{\alpha \beta \gamma}_{\eta}T^{\eta} \nonumber \\&&\langle T^{\alpha},
T^{\beta} \rangle = h^{\alpha\beta} \nonumber \\&& [X^{M},X^{N},X^{L}]=[X^{M}_{\alpha}T^{\alpha},X^{N}_{\beta}
T^{\beta},X^{L}_{\gamma}T^{\gamma}]\nonumber \\&&\langle X^{M},X^{M}\rangle = X^{M}_{\alpha}X^{M}_{\beta}\langle
T^{\alpha}, T^{\beta} \rangle \; .
\label{t8}
\end{eqnarray}
This equation shows that the relevant action of $M0$-branes has a three dimensional Nambu-Poisson bracket
with the Li-$3$-algebra \cite{R20,R21,R22,R23}. To obtain the relevant action for D3-branes (equation (\ref{ttt5})),
we use the following mappings \cite{R5,R14,R15,R16,R17,R18,R19,R20,R21,R22,R23}:

\begin{eqnarray}
&&X^{a}=E^{a}\quad X^{i}=A^{i} \quad X^{j}=r\Theta^{j}\; , \nonumber \\
&&\nonumber \\
&&E^{a}=(-1.a,a,a)\quad E_{a}=(1,a,a,a)\quad F^{ab}=(-1,a^{2},a^{2},a^{2})=g^{ab} \; , \nonumber \\
&&\nonumber \\
&&
E^{a}_{\alpha}E_{b}^{\beta}=\delta_{a}^{b}\delta_{\alpha}^{\beta}\quad
\langle E_{a},E_{b}\rangle=F_{ab}
\quad \langle[E_{a},X^{b},\Theta^{j}],[E^{b},X^{a},\Theta^{i}]\rangle=
\delta_{a}^{b} \partial_{a}\Theta^{i}\partial_{b}\Theta^{j} \; , \nonumber \\
&&\nonumber \\
&&
\langle[E_{a},E^{b},E^{c}],[E^{a},E^{b},E^{c}]\rangle=\delta^{a}_{a}(F^{bc})^{2}\quad \langle[E_{a},X^{b},
\Theta^{j}],[E^{b},X^{a},A^{i}]\rangle=\delta_{a}^{b} 2\pi \alpha' \partial_{a}A^{i}\partial_{b}A^{j} \; , \nonumber \\
&&\nonumber \\ &&
\langle[E_{a},A^{i},\Theta^{j}],[E^{b},A^{j},\Theta^{i}]\rangle=\frac{\delta_{a}^{b}}{2\pi
\alpha'}A^{i}A^{j}\Theta^{j}\Theta^{i} \quad
\langle[E_{a},E_{b},E_{c}],[E^{a},E^{b},E^{c}]\rangle=\delta_{a}^{a}\delta_{b}^{b}\delta_{c}^{c} \; , \nonumber \\
&&\nonumber \\
 &&\Sigma_{m}\rightarrow \frac{1}{(2\pi)^{p}}\int d^{p+1}\sigma
\Sigma_{m-p-1} i,j=p+1,..,10\quad a,b=0,1,...p\quad m,n=0,..,10~~ \; ,
\label{tttt5}
\end{eqnarray}
where $a$ is the scale factor of our universe which is located on the $D3$-brane and $g^{ab}$ is the
metric of the universe and its related $D3$-brane. Also, $E$ is the field which lives on the $D3$ brane
and $A$ and $\Theta$ are fields which live on the $D8$ and $D4$ branes. Summing over the action of the
$M0$-brane in equation (\ref{t7}) and applying mappings in equation (\ref{tttt5}) yields:

\begin{eqnarray}
&&S_{Sum}= \Sigma_{a=0}^{3}S_{M0}=-\Sigma_{a=0}^{3}T_{M0} \int dt~
Tr\left( \Sigma_{m=0}^{9}
\langle[X^{a},X^{b},X^{c}],[X^{a},X^{b},X^{c}]\rangle \right) = \nonumber
\\ && - \frac{2T_{3}}{g_{s}}\int d^{4}x
\left(1+\frac{r^{2}}{4}\Theta'^{2}+2\pi \alpha' \left(\frac{d
A^{i}}{dt}\right)^{2}+\frac{r^{2}}{2\pi
\alpha'}\Theta^{2}(A^{i})^{2}+F_{ab}^{2}\right)=S_{DBI-D3} \; ,
\label{ttttt5}
\end{eqnarray}
where we have made use of the fact that $T_{M0}=\frac{2T_{3}}{(2\pi)^{3}g_{s}}$. This equation shows that the
$D3$-brane is constructed from joining $N$ $M0$-branes. To obtain the explicit form of this action, we
should calculate $\Theta$ and $T$ in terms of time. For this reason, we consider the effect of other
branes by deriving their actions and relations between fields on each brane. The effective tachyonic
action for the $D4$-brane can be written as \cite{R13}:

\begin{eqnarray}
&&S_{D4}=-\frac{2T_{4}}{g_{s}}\int d^{5}\sigma V(T)\sqrt{-det D},\nonumber\\&&
D_{ab}=\Big(g_{MN}-\frac{T^{2}l^{2}}{Q}g_{M4}g_{N4}\Big)\partial_{a}X^{M}\partial_{b}X^{N}
+(F^{MN})^{2}+\frac{1}{2Q}\Big((D_{a}T)(D_{b}T)^{\star}+(D_{a}T)^{\ast}(D_{b}T)\Big)+\nonumber\\
&&il\Big(g_{a4}+\partial_{a}X^{i}g_{44}\Big)\Big(T(D_{b}T)^{\ast}-T^{\ast}(D_{b}T)\Big)+
il\Big(T(D_{a}T)^{\ast}-T^{\ast}(D_{a}T)\Big)\Big(g_{b4}+\partial_{b}X^{i}g_{44}\Big) \; ,\nonumber
\\&&\nonumber
\\&& Q=1+\frac{T^{2}}{2\pi \alpha'}l^{i}l^{j}g_{ij}\quad D_{a}T=\partial_{a}T-i E_{a}T \; ,
\nonumber
\\&& V(T)=\frac{1}{{{}\cosh(\sqrt{\pi T})}} \; ,
\label{tq5}
\end{eqnarray}
where $i,j=5,..,10$ and $a,b=0,1,2,3,4$. When intersecting branes are away from each other, the tachyon
{{} ($T$)} is replaced by $A^{i}$. Using this, and assuming that the fields are small, we can
rewrite the action of the $D4$-brane as:

\begin{eqnarray}
&& V(T)=1 \quad F^{ij}=E^{i}E^{j}, \nonumber\\&&
E_{i}E^{j}=\delta_{i}^{j} \quad X^{M}=E^{M} \quad l_{i}=r
\gamma\Theta, \nonumber\\&& X^{N}=\Theta^{N}\quad T=A^{i}\quad 2\pi
\alpha'=1, \nonumber\\&& g_{MN}=F_{MN}=E_{M}E_{N}\nonumber\\&&
T=T^{\ast}\rightarrow
T(D_{a}T)^{\ast}-T^{\ast}(D_{a}T)=2iE_{a}T^{2}, \nonumber\\&&
\rightarrow T(D_{b}T)^{\ast}-T^{\ast}(D_{b}T)=2iE_{b}T^{2} \; ,
\nonumber\\&& S_{D4} =
 -\frac{2T_{4}}{g_{s}} \int d^{5}\sigma \sqrt{U},\nonumber\\&&
U=1+\partial_{a}F^{ij}\partial_{b}F^{ij}+\partial_{a}A^{j}\partial_{b}A^{i}-
(A^{j})^{2}-(F^{ij})^{2}-4E_{i}
E_{j}\partial_{a}E^{j}\partial_{b}E^{i}+r^{2}\partial_{a}(\Theta^{2})\partial_{b}(\Theta^{2}) \; .
\label{tqq5}
\end{eqnarray}
We can re-obtain the relevant action for $D4$-branes in the background of the $D3$ and anti-$D8$-branes  by
summing over the action of the $M0$-branes and apply the following mappings
\cite{R5,R14,R15,R16,R17,R18,R19,R20,R21,R22,R23}:

\begin{eqnarray}
&&X^{a}=\Theta^{a}\quad X^{i}=E^{i}\quad X^{j}=A^{j}\quad
\langle[X^{a},\Theta^{b},\Theta^{c}],[X^{b},\Theta^{b},\Theta^{c}]\rangle=\partial_{a}\Theta^{2}\partial_{b}\Theta^{2} \; ,
\nonumber \\
&&\nonumber \\
&& E^{i}_{\alpha}E_{j}^{\beta}=\delta_{i}^{j}\delta_{\alpha}^{\beta}\quad \langle E_{i},E_{j}\rangle=F_{ij}
\quad \langle[X^{a},X^{b},F^{ij}],[X^{a},X^{b},F^{ij}]\rangle=\partial_{b}(E_{i}\partial_{a}E^{j})
\partial_{a}(E_{j}\partial_{b}E^{j}) \; , \nonumber \\
&&\nonumber \\
&&
\langle[X^{a},E^{i},E^{j}],[X^{b},E^{i},E^{j}]\rangle=\partial_{a}F^{ij}\partial_{b}F^{ij}\quad \langle[E_{k},E^{i},E^{j}],
[E^{k'},E^{i},E^{j}]\rangle=-\delta_{k}^{k'}F^{ij}F^{ij} \; , \nonumber \\
&&\nonumber \\ &&
\langle[E_{i},X^{a},A^{j}],[E^{j},X^{b},A^{j}]\rangle=\delta_{i}^{j}\partial_{a}A^{j}\partial_{b}A^{i}\quad
 \langle[E_{i},E^{i},A^{j}],[E_{j},E^{j},A^{j}]\rangle=-\delta_{i}^{j}\delta_{i}^{j}A^{j}A^{i} \; , \nonumber \\
&&\nonumber \\
 &&\langle[E_{i},X^{b},E^{j}],[E_{j},X^{b},E^{i}]\rangle=-E_{i}
E_{j}\partial_{a}E^{j}\partial_{b}E^{i}  \quad
  \langle[E_{i},E_{j},E_{k}],[E^{i},E^{j},E^{k}]\rangle=\delta_{i}^{j}\delta_{i}^{j}\delta_{k}^{k}, \nonumber \\
&&\nonumber \\
 &&\Sigma_{m}\rightarrow \frac{1}{(2\pi)^{p}}\int d^{p+1}\sigma
\Sigma_{m-p-1} i,j=p+1,..,10\quad a,b=0,1,...p\quad m,n=0,..,10. ~~
\label{t9}
\end{eqnarray}

Using the above equations, the action of the compacted $D4$-brane in the background of the $D3$ and anti-$D8$
branes can be derived as \cite{R5,R14,R15,R16,R17,R18,R19,R20,R21,R22,R23}:

 \begin{eqnarray}
&& S_{compact-D4} = \Sigma_{a=0}^{4}S_{M0}=-\Sigma_{a=0}^{4}T_{M0}
\int dt Tr( \Sigma_{m=0}^{9}
\langle[X^{a},X^{b},X^{c}],[X^{a},X^{b},X^{c}]\rangle) = \nonumber
\\ && -\frac{2T_{4}}{g_{s}} \int d^{5}\sigma Tr (\Sigma_{a,b,c=0}^{4}
\Sigma_{i,j,k=5}^{10}
\{1+\partial_{a}F^{ij}\partial_{b}F^{ij}+\delta_{i}^{j}\partial_{a}A^{j}\partial_{b}A^{i}+
\partial_{b}(E_{i}\partial_{a}E^{j})\partial_{a}(E_{j}\partial_{b}E^{i})\nonumber\\&&-
\delta_{i}^{j}\delta_{i}^{j}A^{j}A^{i}-\delta_{k}^{k}(F^{ij})^{2}-4E_{i}
E_{j}\partial_{a}E^{j}\partial_{b}E^{i}+r^{2}\partial_{a}\Theta^{2}\partial_{b}\Theta^{2}
\}) \; ,
\label{t10}
\end{eqnarray}
where we have made use of  the fact that $T_{M0}=\frac{2T_{4}}{(2\pi)^{4}g_{s}}$.
This action is the same action as in (\ref{tqq5}) and includes extra terms
 due to the compactification of the $D4$ brane. Now, we assume that our universe is
located on the $D3$-brane and interacts with the $D4$-brane via $F^{ij}$ and
this field plays the role of the graviton for it. Thus, this field has
a direct relation with the metric on $D3$ ($g^{ij}=\eta^{ij}+F^{ij})$
and $E^{i}$ can be written as a function of the scale factor of the
universe ($E^{i}=(-1,a,a,a)$). On the other hand, the coordinates of the
compacted D4 brane are ($X^{0},X^{1},X^{2},X^{3},\theta$). Assuming that
fields are small and only a function of $\theta$, we can write the
action of the compacted $D4$ brane as follows:

\begin{eqnarray}
&& S_{compact-D4} =-T_{D4} \int d^{5}\sigma Tr \left(
\Sigma_{i,j=5}^{10}\left(1+\left(\frac{\partial F^{ij}}{\partial
\theta}\right)^{2}+ \left(\frac{\partial A^{i}}{\partial
\theta}\right)^{2}-
(A^{j})^{2}-(F^{ij})^{2}+\nonumber\right.\right.\\&&  \left.\left.
\left(\frac{\partial}{\partial \theta}(E_{i}\frac{\partial
E^{j}}{\partial \theta})\right)^{2}
-4E_{i}^{2}\left(\frac{\partial E^{i} }{\partial
\theta}\right)^{2} +r^{2}\left(\frac{\partial \Theta^{2}
}{\partial \theta}\right)^{2} \right)\right) \; .
\label{t11}
\end{eqnarray}

The equations of motion obtained from the above equation are:
\begin{eqnarray}
&&  \frac{\partial^{2} F^{ij}}{\partial
\theta^{2}}+F^{ij}=0\Rightarrow F^{ij}=M\sin(\theta)+N
\cos(\theta) \; , \nonumber\\&& \frac{\partial^{2} A^{i}}{\partial
\theta^{2}}+A^{i}=0\Rightarrow A^{i}=M'\sin(\theta)+N'
\cos(\theta)\nonumber\\&& \frac{\partial^{2}}{\partial
\theta^{2}}(E_{i} \frac{\partial E^{j}}{\partial
\theta})+4E_{i}\frac{\partial E^{j}}{\partial \theta}=0\Rightarrow
E_{i}\frac{\partial E^{j}}{\partial \theta}= M'' \sin(2\theta)+N''
\cos (2\theta) \; , \nonumber\\&& r=constant\quad \frac{\partial^{2}
\Theta^{2}}{\partial \theta^{2}}=0 \Rightarrow \frac{\partial
\Theta^{2}}{\partial \theta}=c\Rightarrow
\Theta^{2}=c\theta+c' \; .
\label{t12}
\end{eqnarray}
This equation shows that all fields depend on $\theta$ and
oscillate with decreasing or increasing the angle between
intersecting branes. Now, we obtain the angle as a function of
time by calculating the action of the rotating anti-$M8$ brane and deriving
the equation of motion. To this end, we can use the following action:

\begin{eqnarray}
&&S_{anti-D8}=-\frac{2T_{8}}{g_{s}}\int d^{9}\sigma \sqrt{-det (
g_{MN}\partial_{a}X^{M}\partial_{b}X^{N})-r^{2}\omega^{2}\Theta^{2}}\nonumber\\&&
X^{0}=t \quad X^{M}=r\Theta^{M} \quad X^{N}=A^{N}\quad
g_{MN}=E_{M}E_{N}=F_{MN}\nonumber\\&&\Rightarrow
S_{anti-D8}=-\frac{2T_{8}}{g_{s}}\int d^{9}\sigma \sqrt{1+
r^{2}(\partial_{t}\Theta)^{2}-\omega^{2}r^{2}\Theta^{2}+(\partial_{t}F^{ij})^{2}+(\partial_{t}A^{i})^{2}} \; .
\label{tqp5}
\end{eqnarray}

We can re-obtain the action of the anti-$M8$-brane in the back ground of
other branes by using the following mappings:
\cite{R5,R14,R15,R16,R17,R18,R19,R20,R21,R22,R23}:

\begin{eqnarray}
&&
\langle[X^{0},E^{i},E^{j}],[X^{0},E^{i},E^{j}]\rangle=(\partial_{t}F^{ij})^{2}\quad \langle[A_{k},E^{i},E^{j}],[A_{k'},
E^{i},E^{j}]\rangle=F^{ij}F^{ij}A_{k}A_{k'} \; , \nonumber \\
&&\nonumber \\ &&
\langle[E_{i},X^{a},A^{j}],[E_{j},X^{b},A^{j}]\rangle=\delta_{i}^{j}\partial_{a}A^{j}\partial_{b}A^{i}
\quad
\langle[E^{i},A^{i},A^{i}],[E^{j},A^{j},A^{j}]\rangle=F^{ij}(A^{i})^{4} \; ,
\nonumber \\
&&\nonumber \\
&&
\langle[E_{i},E_{j},E_{k}],[E^{i},E^{j},E^{k}]\rangle=\delta_{i}^{j}\delta_{i}^{j}\delta_{k}^{k}\quad
\langle[E_{i},X^{a},\Theta^{j}],[E^{j},X^{b},\Theta^{j}]\rangle=\delta_{i}^{j}\partial_{a}
\Theta^{j}\partial_{b}\Theta^{i} \; ,\nonumber \\
&&\nonumber \\
&&
\langle[E_{i},E^{i},\Theta^{j}],[E_{j},E^{j},\Theta^{j}]\rangle=-\omega^{2}\delta_{i}^{j}\delta_{i}^{j}\Theta^{j}
\Theta^{i}\quad \Sigma_{m}\rightarrow \frac{1}{(2\pi)^{p}}\int d^{p+1}\sigma
\Sigma_{m-p-1} i,j=p+1,..,10\quad a,b=0,1,...p \; .
\label{t13}
\end{eqnarray}
Using the above mappings, we can obtain the relevant action for the anti-$D8$-brane in the background
of the $D3$ and $D4$ branes:

\begin{eqnarray}
&& S_{anti-D8} = \Sigma_{a=0}^{8}S_{M0}=-\Sigma_{a=0}^{8}T_{M0}
\int dt Tr\left( \Sigma_{m=0}^{9}
\langle[X^{a},X^{b},X^{c}],[X^{a},X^{b},X^{c}]\rangle \right) = \nonumber
\\ && -\frac{2T_{D8}}{g_{s}} \int d^{9}\sigma Tr \Biggl(\Sigma_{a,b,c=0}^{8}
\Sigma_{i,j,k=9}^{10} \Bigl\{1+
r^{2}(\partial_{t}\Theta)^{2}-\omega^{2}r^{2}\Theta^{2}+(\partial_{t}F^{ij})^{2}+(\partial_{t}A^{i})^{2}\nonumber\\&&
+F^{ij}(A^{i})^{4}+(F^{ij})^{2}(A^{i})^{2} \Bigr\}\Biggr) \; ,
\label{t14}
\end{eqnarray}
which is the same as that of action (\ref{tqp5}) in addition to some
extra terms which emerge due to the  interaction between branes.
This action gives the following equations of motion:

\begin{eqnarray}
&& r= constant\rightarrow \frac{\partial^{2} \Theta }{\partial
t^{2}}+\omega^{2}\Theta=0\Rightarrow \Theta=D \cos(\omega t)+
D'\sin(\omega t) \quad And \quad \Theta^{2}=c\theta + c'
\Rightarrow \nonumber\\&&D^{2}+(D')^{2}+D D'\sin(2\omega
t)=c\theta + c' \Rightarrow \nonumber\\&& D=D'=\sqrt{\pi} \quad
c'=2\pi \quad c=\frac{1}{2}\rightarrow \theta=2\pi \sin(2\omega t)
\label{t15}\\&&\nonumber\\&&\nonumber\\&&
V(A)=-(F^{ij}(A^{i})^{4}+(F^{ij})^{2}(A^{i})^{2})\nonumber\\&&
\Rightarrow m^{2}=-\frac{\partial^{2}V}{\partial
A^{2}}=12F^{ij}(A^{i})^{2}+2(F^{ij})^{2}\nonumber\\&&F^{ij}=N\cos(\theta)+M
\sin(\theta) \quad A^{i}=N'\cos(\theta)+M'
\sin(\theta)\nonumber\\&& \frac{1}{c}=2,
M=1,N=0,M'=N'=1\Rightarrow \nonumber\\&&F^{ij}= \sin(2\pi
\sin(2\omega t))\quad A^{i}=\sin(2\pi \sin(2\omega t))+ \cos(2\pi
\sin(2\omega t))\Rightarrow \nonumber\\&& m^{2}= 12\sin(2\pi
\sin(2\omega t))\Big(\sin(2\pi \sin(2\omega t)) + \sin(2\pi
\sin(2\omega t))\Big)^{2} +2\cos^{2}(2\pi \sin(2\omega
t))\nonumber\\&& m^{2}= 12\sin(2\pi \sin(2\omega
t))\Big(1+\sin(4\pi \sin(2\omega t))\Big) +2\sin^{2}(2\pi
\sin(2\omega t))\Rightarrow \nonumber\\&& \theta \leq 0
\Rightarrow \frac{T}{4} \leq t \leq \frac{T}{2} \Rightarrow
m^{2}\leq 0 \Rightarrow A\rightarrow T \nonumber\\&& T \rightarrow
A \Rightarrow \theta=\theta+\pi \Rightarrow
m^{2}\geq 0\label{t16}\\
\nonumber\\&& \nonumber\\&& E^{j}=a(\theta); \quad
E_{j}=a^{-1}(\theta); \quad \theta=2\pi \sin(2\omega t) \rightarrow
d\theta=4\pi \omega \cos(2\omega t)dt\Rightarrow\nonumber\\&&
E_{i}\frac{\partial E^{j}}{\partial \theta}=E_{i}\frac{\partial
E^{j}}{4\pi \omega \cos(2\omega t)\partial t}=\frac{\frac{d
a}{dt}}{4\pi \omega \cos(2\omega t)a}=\frac{H}{4\pi \omega
\cos(2\omega t)}= M'' \sin(2\theta)+N'' \cos (2\theta)\Rightarrow
\nonumber\\&& H=(4\pi \omega \cos(2\omega t))(M'' \sin(2\theta)+N''
\cos (2\theta))\nonumber\\&& \dot{H}=-(8\pi \omega^{2}
\sin(2\omega t))(M'' \sin(2\theta)+N'' \cos
(2\theta))+\dot{\theta}(4\pi\omega \cos(2\omega t))(M''
\cos(2\theta)+N'' \sin(2\theta))\label{t17}
\end{eqnarray}
Equation (\ref{t15}) shows that the angle between intersecting
branes oscillates with time and causes the increase and decrease
in the values of the fields which live on the $D3$, $D4$ and anti-$D8$ branes. As can be
seen from equation (\ref{t16}), with the passage of time and approaching
branes towards each other ($\theta\rightarrow 0$) or moving away
from each other ($\theta\rightarrow \pi$), fields on the
intersecting branes ($A^{i}$) gain negative mass
($m^{2}\rightarrow -m^{2}$) and become tachyons. To remove these
states, intersecting branes rebound, ($\theta\rightarrow \theta
+\pi$) and ($T \rightarrow A^{i}$). It is clear from equation
(\ref{t17}) that the Hubble parameter (H) depends on time and is
positive for $t\leq \frac{T}{4}$ and  negative between
$\frac{T}{4}$ and $\frac{T}{2}$ which is a signature of the
contracting branch of the universe. Substituting the equations
(\ref{t12}, \ref{t15}, \ref{t16}, \ref{t17}) in action (\ref{ttt5}),
we get:
\begin{eqnarray}
&&S_{DBI-D3}=\frac{-2T_{3}}{g_{s}}\int d^{4}x
\left(1+\frac{r^{2}}{4}\Theta'^{2}+2\pi \alpha' \left(\frac{d
A^{i}}{dt}\right)^{2}+\frac{r^{2}}{2\pi
\alpha'}\Theta^{2}(A^{i})^{2}+F_{ab}^{2}\right)=\nonumber
\\&& \frac{-2T_{3}}{g_{s}}\int d^{4}x
\Big(1+\frac{r^{2}}{8} \frac{\theta'^{2}}{\theta}+2\pi \alpha'
\dot{\theta}^{2}(-N'\sin(\theta)+M'\cos(\theta))^{2}+\nonumber
\\&&\frac{r^{2}}{2\pi \alpha'}\theta (N'\cos(\theta)+M'
\sin(\theta))^{2}+(N \cos(\theta)+M
\sin(\theta))^{2}\Big)\nonumber
\\&&\nonumber
\\&& M=1,N=0,M'=N'=1\Rightarrow S_{DBI-D3}=\frac{-2T_{3}}{g_{s}}\int d^{4}x \Big(1+\frac{r^{2}}{8} 
\frac{\theta'^{2}}{\theta}+
2\pi \alpha'
\dot{\theta}^{2}(-\sin(\theta)+\cos(\theta))^{2}+\nonumber
\\&&\frac{r^{2}}{2\pi \alpha'}\theta (\cos(\theta)+
\sin(\theta))^{2}+sin^{2}(\theta)\Big)\Rightarrow \nonumber
\\&&\nonumber
\\&& S_{DBI-D3}=\frac{-2T_{3}}{g_{s}}\int d^{4}x \Big(1+\frac{r^{2}}{8}\frac{\theta'^{2}}{\theta}+
2\pi \alpha' \dot{\theta}^{2}(1-\sin(2\theta))+\frac{r^{2}}{2\pi
\alpha'}\theta (1+\sin(2\theta))+\sin^{2}(\theta)\Big)\nonumber
\\&&\nonumber
\\&& N''=0, M''=1\Rightarrow \sin(2\theta)=\frac{H}{(4\pi\omega \cos(2\omega t))}=
\frac{H}{\lambda^{2}(t)} = 4\pi \omega
\sin(2\omega t)\nonumber
\\&& 2\sin^{2}(\theta)=1-\cos(2\theta)=1-\sqrt{1-\sin^{2}(2\theta)}=
1-\sqrt{1-\frac{H^{2}}{\lambda^{2}(t)}}\Rightarrow \nonumber
\\&&\nonumber
\\&&S_{DBI-D3}=\frac{-2T_{3}}{g_{s}}\int d^{4}x \left(1+\frac{r^{2}}{8}\frac{\theta'^{2}}{\theta}+
2\pi \alpha'
\dot{\theta}^{2}\left(1-\frac{H}{\lambda(t)}\right)+\frac{r^{2}}{2\pi
\alpha'}\theta
\left(1+\frac{H}{\lambda(t)}\right)+\frac{1}{2}\left(1-\sqrt{1-\frac{H^{2}}{\lambda^{2}(t)}}\right)\right)\nonumber
\\&& \nonumber
\\&& \Rightarrow S_{DBI-D3}=S_{torsion}+S_{remain} \nonumber
\\&& \nonumber
\\&&S_{torsion}=\frac{-2T_{3}}{g_{s}}\int d^{4}x \left(\frac{r^{2}}{2\pi
\alpha'}
\frac{H}{\lambda(t)}\theta+\frac{1}{2}\left(1-\sqrt{1-\frac{H^{2}}{\lambda^{2}(t)}}\right)\right)=
\nonumber
\\&&\frac{-2T_{3}}{g_{s}}\int d^{4}x \left(\frac{r^{2}}{2\pi
\alpha'}\frac{H}{\lambda(t)} \arcsin\left(\frac{H}{2\lambda(t)}\right)
+\frac{1}{2}\left(1-\sqrt{1-\frac{H^{2}}{\lambda^{2}(t)}}\right)\right)=\nonumber
\\&&\frac{-2T_{3}}{g_{s}}\int d^{4}x \left(\frac{r^{2}}{2\pi
\alpha'}  \frac{\sqrt{-{\mathcal T}}}{\sqrt{6}\lambda(t)} \arcsin\left(\frac{\sqrt{-{\mathcal T}}}{2\sqrt{6}\lambda(t)}\right)
+\frac{1}{2}\left(1-\sqrt{1+\frac{{\mathcal T}}{6\lambda^{2}(t)}}\right)\right)\nonumber
\\&&\nonumber
\\&& M=1,N=0\Rightarrow L_{tele-LQC}= -\frac{r^{2}}{2\pi
\alpha'}\frac{\sqrt{-{\mathcal T}}}{\sqrt{6}\lambda(t)}\arcsin\left(\frac{\sqrt{-{\mathcal T}}}{2\sqrt{6}\lambda(t)}\right)
+\frac{1}{2}\left(1-\sqrt{1+\frac{{\mathcal T}}{6\lambda^{2}(t)}}\right)\quad
\nonumber
\\&&\nonumber
\\&&M=0,N=1\Rightarrow L_{tele-LQC}= \frac{r^{2}}{2\pi\alpha'} \frac{\sqrt{-{\mathcal T}}}{\sqrt{6}\lambda(t)}
\arcsin\left(\frac{\sqrt{-{\mathcal T}}}{2\sqrt{6}\lambda(t)}\right)
+\frac{1}{2}\left(1+\sqrt{1+\frac{{\mathcal T}}{6\lambda^{2}(t)}}\right)\quad
\label{t18}
\end{eqnarray}
When $\frac{r^{2}}{2\pi\alpha'}=1$, the Lagrangian in Eq. (\ref{t18}) is practically
the same as defined in Eqs. (\ref{t1}) and (\ref{t2}) for teleparallel Loop Quantum
Cosmology (LQC). The difference is that, in that case, the critical density given by
$12\lambda^2(t)$ is time dependent. In fact, we find a new concept for the Ashtekar connection
and fields that have been introduced in that model. These results
show that the reason for torsion occurring  is the wrapping and
opening of the $D3$ brane  around the $D4$ brane. Also, the interaction between fields on the
brane causes the change of angle between the intersecting branes and
oscillating of this angle leads to LQC.
\section{The origin of torsion in Einstein cosmology  }\label{o2}
So far, we have proposed a new model which allows us to
consider the origin of teleparallel LQC in a
system of oscillating branes. Now, we show that by ignoring the
effect of other branes, our model produces usual teleparallel
theory in Einstein cosmology. This type of cosmology is obtained
from the following Lagrangian in teleparallel theory \cite{R3}:
\begin{eqnarray}
&& L=\frac{1}{2}{\mathcal T} V -\rho V\label{t19} \; ,
\end{eqnarray}
where
\begin{eqnarray}
&&
{\mathcal T}=S_{\gamma}\smallskip^{\mu\nu}{\mathcal T}^{\gamma}\smallskip_{\mu\nu}\nonumber \; ,
\\&&{\mathcal T}^{\gamma}\smallskip_{\mu\nu}=\Gamma^{\gamma}\smallskip_{\mu\nu}-\Gamma^{\gamma}\smallskip_{\nu\mu} \; ,
\nonumber\\&&
K^{\mu\nu}\smallskip_{\gamma}=-\frac{1}{2}({\mathcal T}^{\mu\nu}\smallskip_{\gamma}-{\mathcal T}^{\nu\mu}\smallskip_{\gamma}-{\mathcal T}_{\gamma}
\smallskip^{\mu\nu})\ \; , \nonumber\\&&
S_{\gamma}\smallskip ^{\mu\nu}=\frac{1}{2}(K^{\mu\nu}\smallskip_{\gamma}-\delta^{\mu}_{\gamma}{\mathcal T}^{\theta\nu}
\smallskip_{\theta}-\delta^{\nu}_{\gamma}{\mathcal T}^{\theta\mu}\smallskip_{\theta}) \; , \nonumber\\&&
\Gamma^{\gamma}\smallskip_{\mu\nu}=E^{\gamma}_{\alpha}\partial_{\nu}E_{\mu}^{\alpha} \; .
\label{t20}
\end{eqnarray}
One can show that the above Lagrangian can be obtained from the following
action in string theory:
\begin{eqnarray}
&&S_{D3}=-\frac{2T_{3}}{g_{s}}\int d^{4}x\sqrt{-det D} \; , \nonumber\\
&& D_{ab}=g_{MN}\partial_{a}X^{M}\partial_{b}X^{N}+
\frac{1}{2}F_{\mu\nu}(\frac{3}{2}F^{\mu\nu}-F^{\theta\nu}-F^{\theta\nu}) \; ,
\nonumber\\&&F_{\mu\nu}=\partial_{\mu}E_{\nu}-\partial_{\nu}E_{\mu} \; .
\label{qq5}
\end{eqnarray}
Now, we obtain the relation between torsion and field strength and
substitute it in the above action to derive the Lagrangian in equation
(\ref{t19}):
\begin{eqnarray}
&&E^{\gamma}F_{\mu\nu}=E^{\gamma}\partial_{\mu}E_{\nu}-E^{\gamma}\partial_{\nu}E_{\mu}\nonumber
\\&&=\Gamma^{\gamma}\smallskip_{\mu\nu}-\Gamma^{\gamma}\smallskip_{\nu\mu}=
{\mathcal T}^{\gamma}\smallskip_{\mu\nu}\nonumber
\\&&E^{\gamma}E_{\gamma}=1\Rightarrow F_{\mu\nu}F^{\mu\nu}=F_{\mu\nu}E^{\gamma}E_{\gamma}F^{\mu\nu}=
{\mathcal T}_{\mu\nu}\smallskip^{\gamma}{\mathcal T}_{\gamma}\smallskip^{\mu\nu}\nonumber
\\&&F_{\mu\nu}F^{\theta\nu}=F_{\mu\nu}E^{\gamma}E_{\gamma}E^{\theta}E_{\theta}F^{\theta\nu}=
{\mathcal T}_{\mu\nu}\smallskip^{\gamma}\delta_{\gamma}^{\theta}{\mathcal T}_{\theta}\smallskip^{\theta\nu}\nonumber
\\&&F_{\mu\nu}\left(\frac{3}{2}F^{\mu\nu}-F^{\theta\nu}-F^{\theta\nu}\right)=F_{\mu\nu}E^{\gamma}E_{\gamma}\left(\frac{3}{2}
F^{\mu\nu}-F^{\theta\nu}-F^{\theta\nu}\right)=\nonumber
\\&&\nonumber
\\&&F_{\mu\nu}E^{\gamma}\Biggl(\frac{1}{2}[E_{\gamma}F^{\mu\nu}-E_{\gamma}F^{\nu\mu}+E_{\gamma}F^{\mu\nu}]-
E_{\gamma}E^{\theta}E_{\theta}F^{\theta\nu}-E_{\gamma}E^{\theta}E_{\theta}F^{\theta\nu}\Biggr)=\nonumber
\\&&{\mathcal T}_{\mu\nu}\smallskip^{\gamma}\Biggl(\frac{1}{2}[{\mathcal T}_{\gamma}\smallskip^{\mu\nu}-{\mathcal T}_{\gamma}\smallskip^{\nu\mu}-
{\mathcal T}^{\mu\nu}\smallskip_{\gamma}]-\delta_{\gamma}^{\theta}{\mathcal T}_{\theta}\smallskip^{\theta\mu}-\delta_{\gamma}^{\theta}
{\mathcal T}_{\theta}\smallskip^{\theta\nu}\Biggr)\nonumber
\\&&={\mathcal T}_{\mu\nu}\smallskip^{\gamma}(K^{\mu\nu}\smallskip_{\gamma}-\delta_{\gamma}^{\theta}
{\mathcal T}_{\theta}\smallskip^{\theta\mu}-\delta_{\gamma}^{\theta}{\mathcal T}_{\theta}\smallskip^{\theta\nu})=
2S_{\gamma}\smallskip^{\mu\nu}{\mathcal T}^{\gamma}\smallskip_{\mu\nu}=2{\mathcal T} \nonumber
\\&&\nonumber
\\&&X^{0}=t\Rightarrow
g_{\mu\nu}\partial_{a}X^{\mu}\partial_{b}X^{\nu}=1+g_{LK}\partial_{a}X^{L}\partial_{b}X^{K},\quad
L,K=1,2,3 \nonumber
\\&&\Rightarrow D_{ab}=1-g_{LK}\partial_{a}X^{L}\partial_{b}X^{K}+
\frac{1}{2}F_{\mu\nu}\left(\frac{3}{2}F^{\mu\nu}-F^{\theta\nu}-F^{\theta\nu}\right)\nonumber
\\&&\Rightarrow S_{D3}=-\frac{2T_{3}}{g_{s}}\int d^{4}x\Biggl[1-\frac{1}{2}g_{LK}\partial_{a}X^{L}\partial_{b}X^{K}+
\frac{1}{4}F_{\mu\nu}\Biggl(\frac{3}{2}F^{\mu\nu}-F^{\theta\nu}-F^{\theta\nu}\Biggr)\Biggr]\nonumber
\\&&\Rightarrow S_{D3}=-\frac{2T_{3}}{g_{s}}\int d^{4}x\left[1-\frac{1}{2}g_{LK}\partial_{a}X^{L}\partial_{b}X^{K}+
\frac{1}{2}{\mathcal T}\right]\nonumber
\\&& \rho= \frac{1}{2}g_{LK}\partial_{a}X^{L}\partial_{b}X^{K}\Rightarrow S_{D3}=-\frac{2T_{3}}{g_{s}}\int d^{4}x\left[1-\rho+
\frac{1}{2}{\mathcal T}\right]\Rightarrow \nonumber
\\&&S_{D3}=-\frac{2T_{3}}{g_{s}}\int dt\smallskip L\Rightarrow L=-\rho \smallskip V+
\frac{1}{2}{\mathcal T}\smallskip V\label{qqq5}
\end{eqnarray}

As can be seen from this equation, the Lagrangian of torsion can
be obtained from the relevant action of $D3$-branes in string theory.
Now, the question arises as to what is the origin of this torsion
in Einstein cosmology? In the previous section, we have shown that the
action of the $D3$ brane can be constructed by summing over the actions of
$M0$-branes. Ignoring fields of other branes, we can re-obtain the
Lagrangian in (\ref{t19}). We use mappings in (\ref{tttt5}) and
the method in \cite{R5,R19,R20,R21,R22,R23} and write:

\begin{eqnarray}
&&X^{\gamma}=E^{\gamma} \quad X_{\mu}=E_{\mu}\quad
[X^{\gamma},X_{\mu},X_{\nu}]=E^{\gamma}\partial_{\mu}E_{\nu}=\Gamma^{\gamma}\smallskip_{\mu\nu}
\nonumber
\\ &&
[X^{\gamma},X_{\nu},X_{\mu}]=-E^{\gamma}\partial_{\nu}E_{\mu}=-\Gamma^{\gamma}\smallskip_{\nu\mu}
\quad [X^{\gamma},X_{\nu},X_{\mu}]=-[X^{\gamma},X_{\mu},X_{\nu}]
\nonumber
\\ &&\nonumber
\\ &&
\Rightarrow \Sigma_{m,n=\mu,\nu}[X^{\gamma},X_{m},X_{n}]=
[X^{\gamma},X_{\mu},X_{\nu}]+[X^{\gamma},X_{\nu},X_{\mu}]
=E^{\gamma}\partial_{\mu}E_{\nu}-E^{\gamma}\partial_{\nu}E_{\mu}=\nonumber
\\ &&\Gamma^{\gamma}\smallskip_{\mu\nu}-\Gamma^{\gamma}\smallskip_{\nu\mu}={\mathcal T}^{\gamma}\smallskip_{\mu\nu}\nonumber
\\ &&  \nonumber
\\ && \Rightarrow\Sigma_{m,n=\mu,\theta}[X^{\theta},X_{m},X_{n}]=
[X^{\theta},X_{\mu},X_{\theta}]+[X^{\theta},X_{\theta},X_{\mu}]
=E^{\theta}\partial_{\mu}E_{\theta}-E^{\theta}\partial_{\theta}E_{\mu}=\nonumber
\\ &&\Gamma^{\theta}\smallskip_{\mu\theta}-\Gamma^{\theta}\smallskip_{\theta\mu}={\mathcal T}^{\theta}\smallskip_{\mu\theta}
\nonumber
\\ &&\nonumber
\\ &&
\Rightarrow\Sigma_{m,n,l=\mu,\nu,\gamma}[X^{m},X^{n},X_{l}]=\frac{1}{2}\Bigl[\Sigma_{m,n=\mu,\nu}\left([X^{m},X^{n},
X_{\gamma}]+[X^{m},X^{n},X_{\gamma}]\right)\Bigr]=\nonumber
\\
&&\frac{1}{2}\Bigl[\Sigma_{m,n=\mu,\nu}\left([X^{m},X^{n},X_{\gamma}]-[X_{\gamma},X^{m},X^{n}]\right)\Bigr]=
\frac{1}{2}\Biggl[\Sigma_{m,n=\mu,\nu}\Bigl([X^{m},X^{n},X_{\gamma}]-\nonumber
\\
&&[X^{n},X^{m},X_{\gamma}]\Bigr)-\Sigma_{m,n=\mu,\nu}[X_{\gamma},X^{m},X^{n}]\Biggr]=
\frac{1}{2}[{\mathcal T}^{\mu\nu}\smallskip_{\gamma}-{\mathcal T}^{\nu\mu}\smallskip_{\gamma}-{\mathcal T}_{\gamma}\smallskip^{\mu\nu}]=
K^{\mu\nu}\smallskip_{\gamma}
\nonumber
\\ &&  \nonumber
\\
&&\Rightarrow\Sigma_{m,n,l=\mu,\nu,\gamma,\theta}[X^{m},X^{n},X_{l}]=\Sigma_{m,n,l=\mu,\nu,\gamma}[X^{m},X^{n},X_{l}]+
\Sigma_{m,l=\mu,\theta}[X^{m},X^{\theta},X_{l}]+\nonumber
\\
&&\Sigma_{n,l=\nu,\theta}[X^{m},X^{\theta},X_{l}]=
K^{\mu\nu}\smallskip_{\gamma}-\Sigma_{m,l=\mu,\theta}[X^{\theta},X^{m},X_{l}]-\Sigma_{n,l=
\mu,\theta}[X^{\theta},X^{n},X_{l}]=\nonumber
\\ &&K^{\mu\nu}\smallskip_{\gamma}- \delta_{\gamma}^{\nu}{\mathcal T}^{\theta\mu}\smallskip_{\theta}-\delta_{\gamma}^{\mu}
{\mathcal T}^{\theta\nu}\smallskip_{\theta}=S^{\mu\nu}\smallskip_{\gamma}\nonumber
\\ &&  \nonumber
\\ && \Rightarrow S_{torsion}= \Sigma_{a=0}^{3}S_{M0}=-\Sigma_{a=0}^{3}T_{M0} \int dt
Tr\Biggl( \Sigma_{l=\gamma,\theta}
\langle\Sigma_{m,n=\mu,\nu,\theta}[X^{m},X^{n},X_{l}],\Sigma_{m,n=\mu,\nu}[X_{m},X_{n},X^{l}]\rangle\Biggr)
= \nonumber
\\ &&\frac{-2T_{3}}{g_{s}}\int d^{4}x
\frac{1}{2}\Bigl(S^{\mu\nu}\smallskip_{\gamma}{\mathcal T}^{\gamma}\smallskip_{\mu\nu}\Bigr)=\frac{-2T_{3}}{g_{s}}\int
d^{4}x \frac{{\mathcal T}}{2}=\frac{-2T_{3}}{g_{s}}\int dt\frac{1}{2}{\mathcal T}
\smallskip V=\frac{-2T_{3}}{g_{s}}\int dt L_{torsion}\Rightarrow \nonumber
\\
&& L_{torsion}= \frac{1}{2}{\mathcal T}
\smallskip V\label{t21}
\end{eqnarray}
This equation shows that by  ignoring the interaction of the $D3$ brane with
intersecting anti-$D8$ and $D4$ branes, the Lagrangian of torsion in
Einstein cosmology can be produced. In fact, this equation implies
that this torsion can be generated as due to the interaction of
scalars in the transverse direction with our own brane. However, loop
quantum cosmology may appear in a system of oscillating
branes in which the angle between intersecting branes changes with time.
\section{Summary and Discussion} \label{sum}
In this research, we have considered the origin of teleparallel
LQC in a system of intersecting branes in which
the angle between branes change with time. This system has been
constructed by two intersecting $D8$-branes, one compacted
$D4$-brane and another $D3$-brane. The $D4$ brane is compacted on a circle and is
located between anti-$D8$-branes. The $D3$ brane wraps around the $D4$ one from one end
and attaches to one of the anti-$D8$-branes from another end. On
each brane, one type of field lives, which interacts with
other fields on  another brane and causes the oscillation of
the angle between the intersecting branes. Decreasing this angle and
approaching the anti-$D8$ branes, the $D3$ brane wraps around the $D4$ one and the contraction
branch begins. Increasing the angle between branes and moving
away from each other, the $D3$ brane separates  from the $D4$ brane and the expansion branch
starts. This wrapping  of the $D3$ brane around the compacted $D4$ one and the opening
of the $D3$ from the $D4$ brane leads to the emergence of teleparallel LQC. Ignoring
the interaction of the $D3$ brane with the intersecting anti-$D8$-branes and $D4$ branes,
and considering the interaction of scalars in the transverse direction with the $D3$ brane,
the Lagrangian of teleparallel LQC is reduced to the Lagrangian of torsion
in Einstein cosmology.
\section*{Acknowledgments}
\noindent  A. Sepehri and A. Pradhan would also like to thank the University of Zululand, 
Kwa-Dlangezwa 3886, South Africa for providing facilities and support where part of this 
work has been done. A. Pradhan also acknowledges IUCAA, Pune, India for awarding a Visiting 
Associateship. The investigation of J. de Haro has been supported in part by MINECO (Spain), 
project MTM2014-52402-C3-1-P. The authors are extremely grateful to one of the referees for 
his valuable suggestions.

\end{document}